\documentclass[3p,twocolumn]{elsarticle}

\usepackage{amsmath}
\DeclareMathOperator*{\Min}{min}

\usepackage{graphicx}
\usepackage{graphicx,epsf} 
\usepackage{dcolumn}
\usepackage{bm}
\usepackage{latexsym}

\begin{document}

\title{Comparative Performance of Tabu Search and Simulated Annealing
  Heuristics for the Quadratic Assignment Problem}

\author{Gerald Paul}
\ead{gerryp@bu.edu}

\address{Center for Polymer Studies and Dept. of Physics, Boston
             University, Boston, MA 02215, USA}

\begin{abstract}

  For almost two decades the question of whether tabu search (TS) or
  simulated annealing (SA) performs better for the quadratic
  assignment problem has been unresolved.  To answer this question
  satisfactorily, we compare performance at various values of targeted
  solution quality, running each heuristic at its optimal number of iterations for each target.  We find that for a number of varied
  problem instances, SA performs better for higher quality targets while 
  TS performs better for lower quality targets.
  
\end{abstract}


\maketitle

\section{Introduction}

The quadratic assignment problem (QAP) is a combinatorial optimization
problem first introduced by Koopmans and Beckman \cite{Koopmans}.  It
is NP-hard and is considered to be one of the most difficult problems
to be solved optimally.  The problem is defined in the following
context: A set of $N$ facilities are to be located at $N$ locations.
The distance between locations $i$ and $j$ is $D_{i,j}$ and the
quantity of materials which flow between facilities $i$ and $j$ is
$F_{i,j}$.  The problem is to assign to each location a single
facility so as to minimize the cost
\begin{equation}C=\sum_{i=1}^N \sum_{j=1}^N  F_{i,j} D_{p(i),p(j)}. \nonumber
\end{equation}
where $p(i)$ represents the location to which facility $i$ is assigned.

There is an extensive literature which addresses the QAP and is
reviewed in \cite{Pardalos,Cela, Anstreicher,James,Loiola}.  With the
exception of specially constructed cases, optimal algorithms have
solved only relatively small instances ($N \le 36$).  Various
heuristic approaches have been developed and applied to problems
typically of size $N\approx 100$ or less. Two of the most successful
heuristics to date for the QAP are tabu search (TS) and simulated
annealing (SA).  They are basic heuristics which are used
alone or as components in hybrid and iterative metaheuristics.

Comparisons of the performance of SA and TS for the QAP have been
inconclusive.  In this work, we are able to successfully characterize
the relative performance of these heuristics by performing the
comparisons for various values of solution quality and by setting the 
number of iterations for each heuristic to the optimal one for the target
solution quality.  

As is common practice, we define the quality, $Q$ of a solution  
\begin{equation}
Q\equiv\frac{C-C_{best}}{C_{best}},  \nonumber
\end{equation}
where $C$ is the value of the objective function for the solution and
$C_{best}$ is the best known value of the objective function for the
instance.  The lower the value of $Q$, the higher the quality.

We find that for each problem instance, there is a value of $Q$, $Q^*$,
above which (lower quality) tabu search performs better --- requires less
time --- than simulated annealing and below which (higher quality)
simulated annealing performs better.

\section{Background}

The tabu search heuristic for the quadratic assignment problem consists of
repeatedly swapping locations of two nodes.  A single iteration of the
heuristic consists of making the swap which most decreases the total
cost.  Under certain conditions, if a move which lowers the cost is
not available, a move which raises the cost is made. To ensure that cycles of
the same moves are avoided, the same move is forbidden ({\it taboo})
until a specified later iteration; we call this later iteration the
{\it eligible iteration} for a given move.  This eligible iteration is
traditionally stored in a {\it tabu list} or {\it tabu table}.  The
process is repeated for a specified number of iterations.

The simulating annealing heuristic also consists of swapping locations
of two facilities.  In the simulated annealing approach used here
\cite{Connolly}, each possible swap is considered in turn and $\delta$, the change in cost for the 
potential swap, is calculated.  The swap is made if $\delta$ is negative 
or if 
\begin{equation}
e^{-\delta/T} > r, \nonumber
\end{equation}
where $T$ is an analog of temperature in physical 
systems that is slowly decreased according to a specified {\it cooling
  schedule} after each iteration and $r$ is a uniformly distributed random variable between 0 and 1.  Randomly making moves 
which increase the cost is done to help escape from local minima.

Pardalos \cite{Pardalos} compared the performance of four
algorithms including simulated annealing and tabu search and found
that ``all of these approaches have almost the same
performance''.  Paulli \cite{Paulli} compared simulated annealing
and tabu search and found that ``when CPU time is taken into
consideration, simulated annealing is clearly preferable to tabu
search''.  On the other hand, \cite{battiti} finds that ``RTS
(Reactive Tabu Search) needs less CPU time than SA to reach average
results in the $1\%$ [of the best known value] region".
In 1998, summarizing the situation, Cela \cite{Cela} commented that
``There is no general agreement concerning the comparison of the
performance  of simulated annealing approaches with that of tabu search approaches for the QAP''.
We are not aware of any later work which has clarified the issue.

\section{Approach}

We address the question of whether tabu search or simulated annealing
performs better for the quadratic assignment problem by recognizing
that the answer depends on desired solution quality and by:
\begin{itemize}
\item defining a performance metric that ensures a fair comparison of different heuristics, 

\item determining the optimal number of iterations for a given target
  quality for TS and SA for each problem instance; for a fair
  comparison of heuristics,  it is critical to run each heuristic
    at its optimal number of iterations for a given target solution
    quality.

\item measuring the performance of TS and SA at multiple target qualities.

\end{itemize}

\section{Performance Metric}

To fairly compare heuristics, solution quality and time must be taken
into account.  Simulated annealing and tabu search are {\it
  multi-start} heuristics; many runs of the heuristic are executed,
each with a different random starting configuration.  A commonly used
performance metric for multi-start heuristics is the percentage of
these runs which attain a specified value of the quality $Q$
(typically $0.01$).  However, this metric doesn't take run time into
account.  Sometimes, the runs times for individual runs of the
heuristics are constrained to be equal but this is problematic
because, as we show below, for a fair comparison each heuristic should
be run at the optimal number of iterations for the quality goal $Q$.
One method of characterizing the performance of multistart heuristics
with different run times employs {\it run-time distributions} of the
times needed across multiple runs to achieve a certain quality goal
(see e.g.  \cite{stut,aiex}).  Instead of using distributions, we
define the performance metric $\bar T(Q,I)$ as the {\it average} time
to attain a quality goal of $Q$ during a set of runs, each run with
$I$ iterations:
\begin{equation}
\bar T(Q,I)= \frac{ \sum_i t_i}{N(Q,I)}, \nonumber
\end{equation}
where $t_i$ is the CPU time for run $i$ and $N(Q,I)$ is the number of
runs which attain a quality goal of $Q$ or better.

Because one heuristic may perform better depending on the quality
goal, we calculate this performance metric not just for a single
quality goal (e.g. $0.01$) but for a range of quality goals.

\section{Numerical Results}

We use C++ implementations of SA and TS in the public domain to
perform our computational experiments.  Both implementations are by
Taillard, and are available at http://mistic.heig-vd\ .ch/taillard/.
The TS code implements the robust tabu search of \cite{Taillard}; the
SA code implements the simulated annealing heuristic of
\cite{Connolly}.  Both implementations are straightforward and a few
pages each in length.  We run the TS heuristic with parameter settings
as described in \cite{Taillard}): tabu list size between $0.9 N$ and
$1.1 N$ and aspiration function parameter equal to $2N^2$; there are no
settable parameters for the SA implementation.

\subsection{Determination of Optimal Number of Iterations}

Given a fixed time in which a heuristic can be executed, there is a tradeoff between the number of iterations per run and the number of runs which can be performed.  The optimal number of iterations per run to reach a quality goal of $Q$, $I_{opt}(Q)$, is the value of $I$ which minimizes $\bar T(Q,I)$.    
We determine $I_{opt}(Q))$ as follows:  For various values of $I$, $I_i$, we run each heuristic multiple times and calculate $\bar T(Q,I_i)$.  Then,
\begin{equation}
I_{opt}(Q) = \{I_i | \bar T(Q,I_i)= \bar T(Q) \}  \nonumber
\end{equation}
where
%
%
\begin{equation}
\bar T(Q) \equiv  \Min_i \bar T(Q,I_i).  \nonumber
\end{equation}
Thus $\bar T(Q)$ is the value of the performance metric when the heuristic is run at $I_{opt}(Q)$ iterations.  

\begin{figure}[tbh]
\centerline{
\epsfxsize=7.0cm
\epsfclipon
\epsfbox{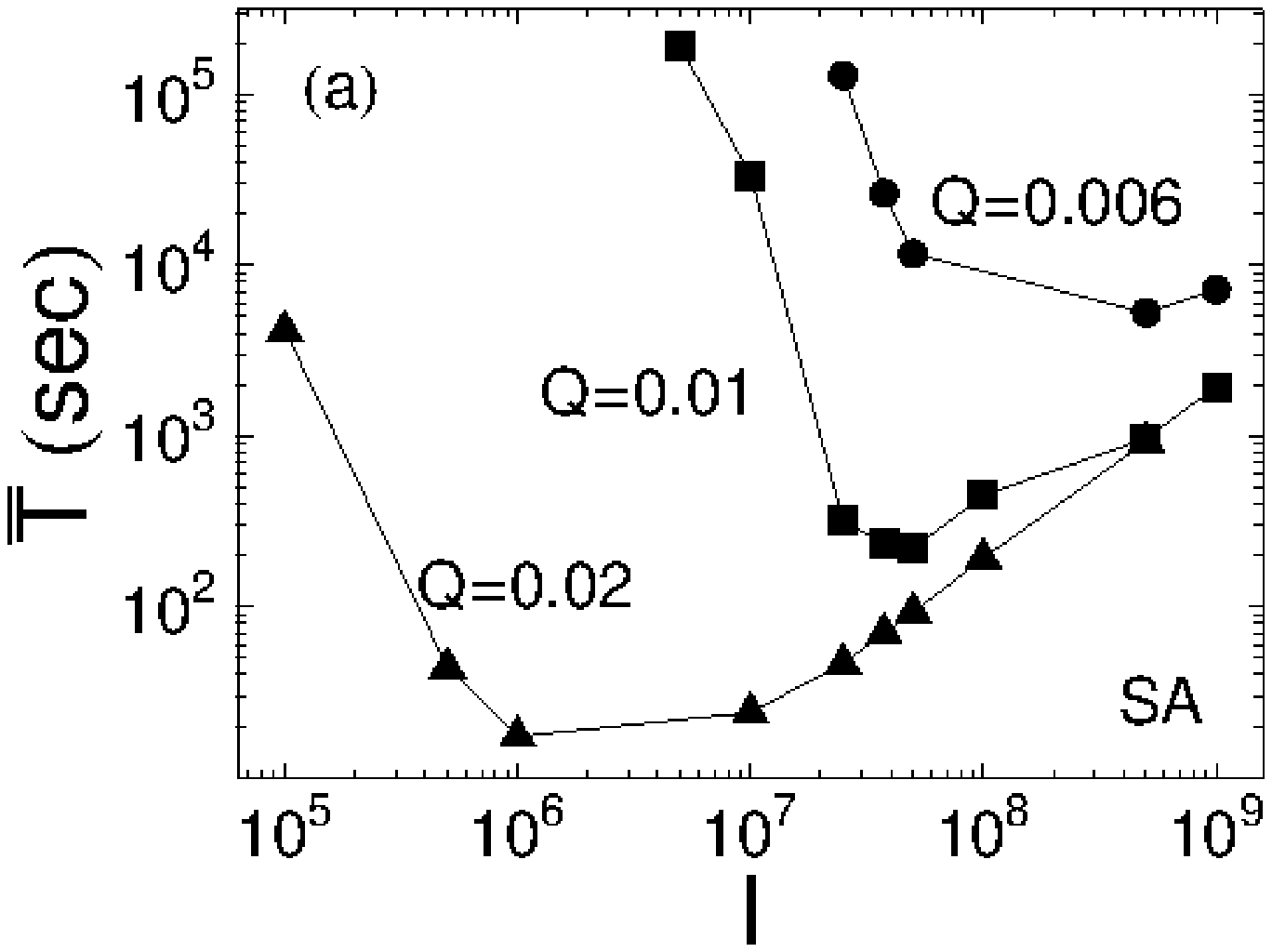}
}
\centerline{
\epsfxsize=7.0cm
\epsfclipon
\epsfbox{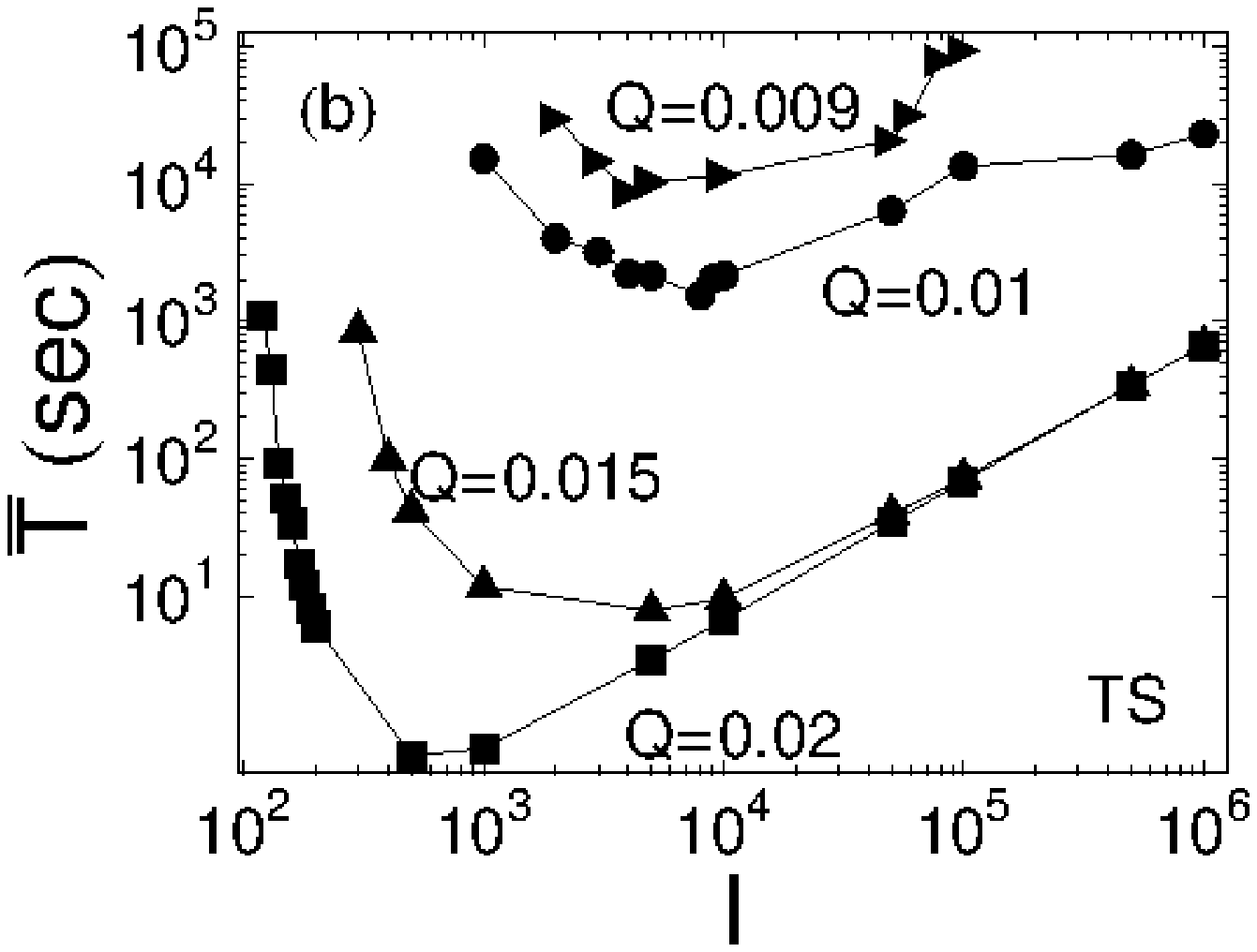}
}
\centerline{
\epsfxsize=7.0cm
\epsfclipon
\epsfbox{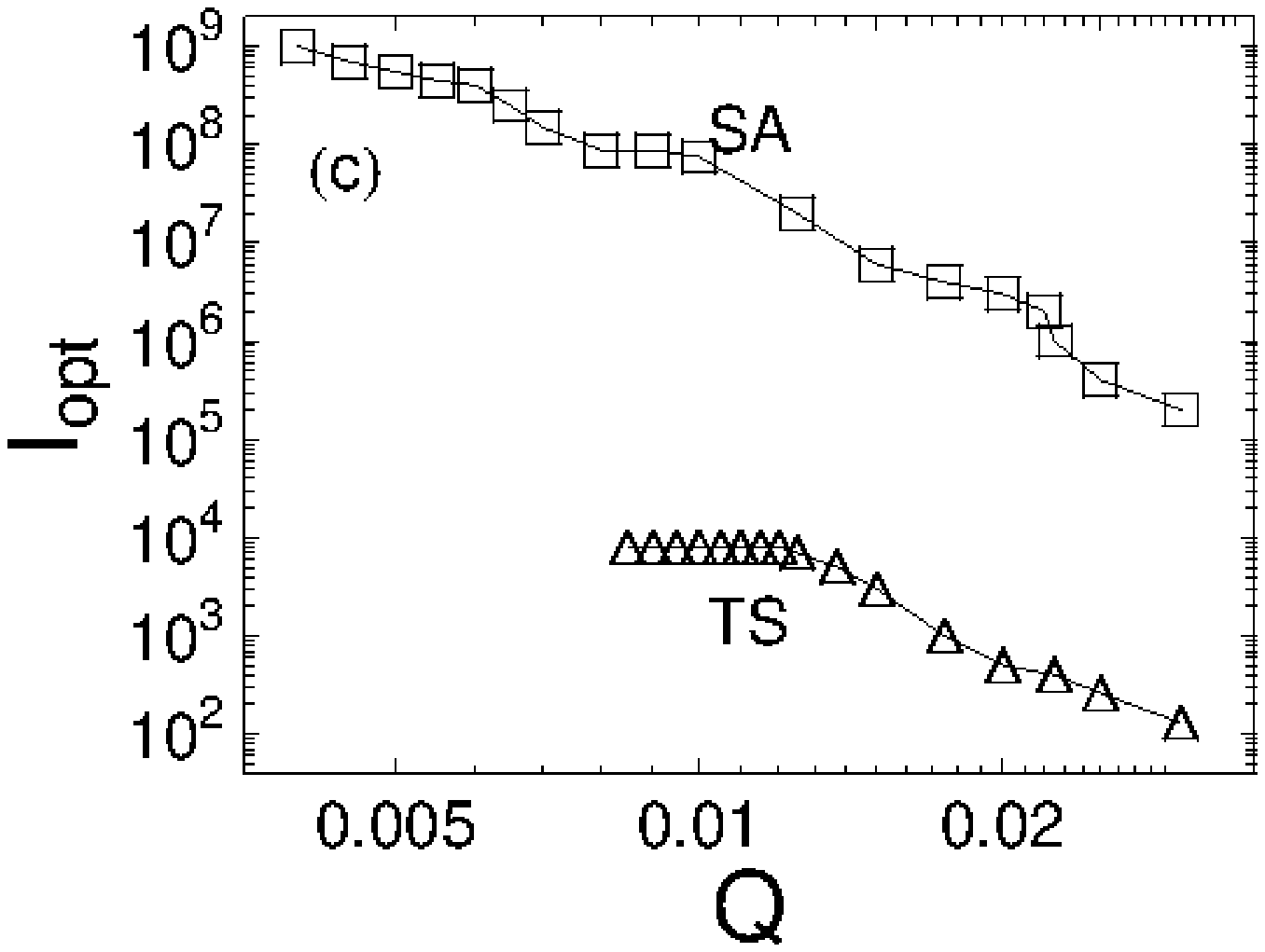}
}
\caption{(a)Dependence of $\bar T$ on number of simulated annealing
  iterations $I$ for quality $Q=0.02,0.01$ and $0.006$ for the QAP
  instance Tai100a.  (b)Dependence of $\bar T$ on number of tabu
  iterations $I$ for quality $Q=0.02, .015, 0.01$ and $0.009$ for the
  QAP instance Tai100a. (c) Optimal number of iterations, $I_{opt}$
  versus $Q$ for SA(squares) and TS (triangles). }
\label{pI}
\end{figure}

In Fig. \ref{pI}(a), using the Tai100a problem instance from QAPLIB \cite{qaplib} as an example, 
we illustrate the process of finding the optimal
number of simulated annealing iterations for
$Q=0.02, 0.01,$ and $0.006$.  The optimal number of iterations,
$I_{opt}$, for each value of $Q$ is the well defined minimum value of
$\bar T$ for each plot.  For a given value of $Q$, we note the large
variation in $\bar T$.  We also note the large variation in $I_{opt}$
for the different values of $Q$.  Thus, choosing a non-optimal value
of iterations (e.g. a single value for the number of iterations for
different $Q$) will result in an unfair characterization of the
performance of the heuristic.  Similarly, Fig. \ref{pI}(b),
illustrates the process of finding the optimal number of tabu
iterations for the instance Tai100a for $Q=0.02, 0.015, 0.01$ and
$0.009$.

In Fig. \ref{pI}(c), we plot $I_{opt}$, versus Q for SA and TS.  For TS, 
$I_{opt}$ increases as $Q$ decreases but 
does not increase below $Q \approx 0.01$.  We infer that for TS there
is no benefit to increasing the number of iterations below this point;
any improvements in quality are gained by running more random starting
configurations. On the other hand, SA benefits by increasing the
number of iterations as $Q$ is decreased over the complete range of $Q$ studied.  
The subject of an optimal number of iterations for the quality goal $Q=0$ for simulated
annealing is treated analytically in \cite{azen}.

\subsection{Performance Comparison of SA and TS}

We perform computational experiments on the following problem instances from
QAPLIB \cite{qaplib} representing a range of problem difficulty, type and size.

\begin{itemize}
\item Tai100a \cite{Taillard} is a totally unstructured instance consisting of random distance and flow matrices.

\item nug30 \cite{nug}, sko100a \cite{sko}, and tho150 \cite{tho} are instances in which the distances are the Manhattan distances 
between locations on a grid.

\item lipa90a \cite{lipa} is a generated problem instance with a known optimal solution.

\item dre110 \cite{dre} is a structured instance consisting of a "grid"
flow matrix with non-zero entries for nearest neighbors only.  It is
part of a series of instances that are specifically designed to be
difficult for heuristics. 

\end{itemize} 

For each problem instance we execute a series of runs for various
values of $I$ including: $I$ = $10^6$, $5 \cdot 10^6$, $10\cdot 10^6$,
$50 \cdot 10^6$, $100 \cdot 10^6$, $500 \cdot 10^6$, and $10^9$ for SA
and $I$ = $10^3$, $5 \cdot 10^3$, $10 \cdot 10^3$, $50 \cdot 10^3$,
$100 \cdot 10^3$, $500 \cdot10^3$, and $10^6$ for TS.  For some
instances, additional values of $I$ are used.  As described above, for
each problem instance we determine the value of $\bar T(Q)$ for
various values of $Q$.

\begin{figure}[h]
\begin{center}$
\begin{array}{cc}
\epsfxsize=7.0cm
\epsfbox{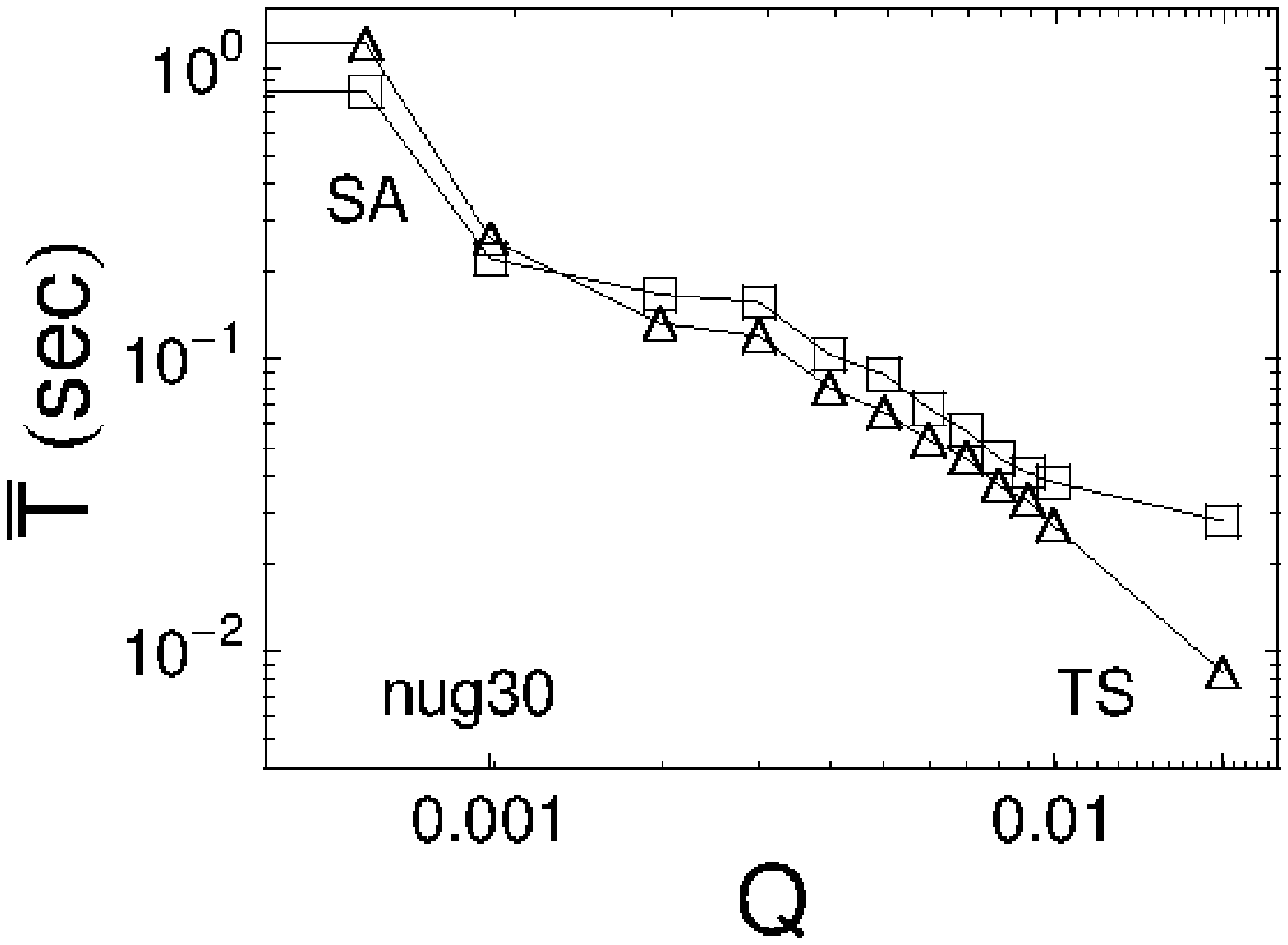} &
\epsfxsize=7.0cm
\epsfbox{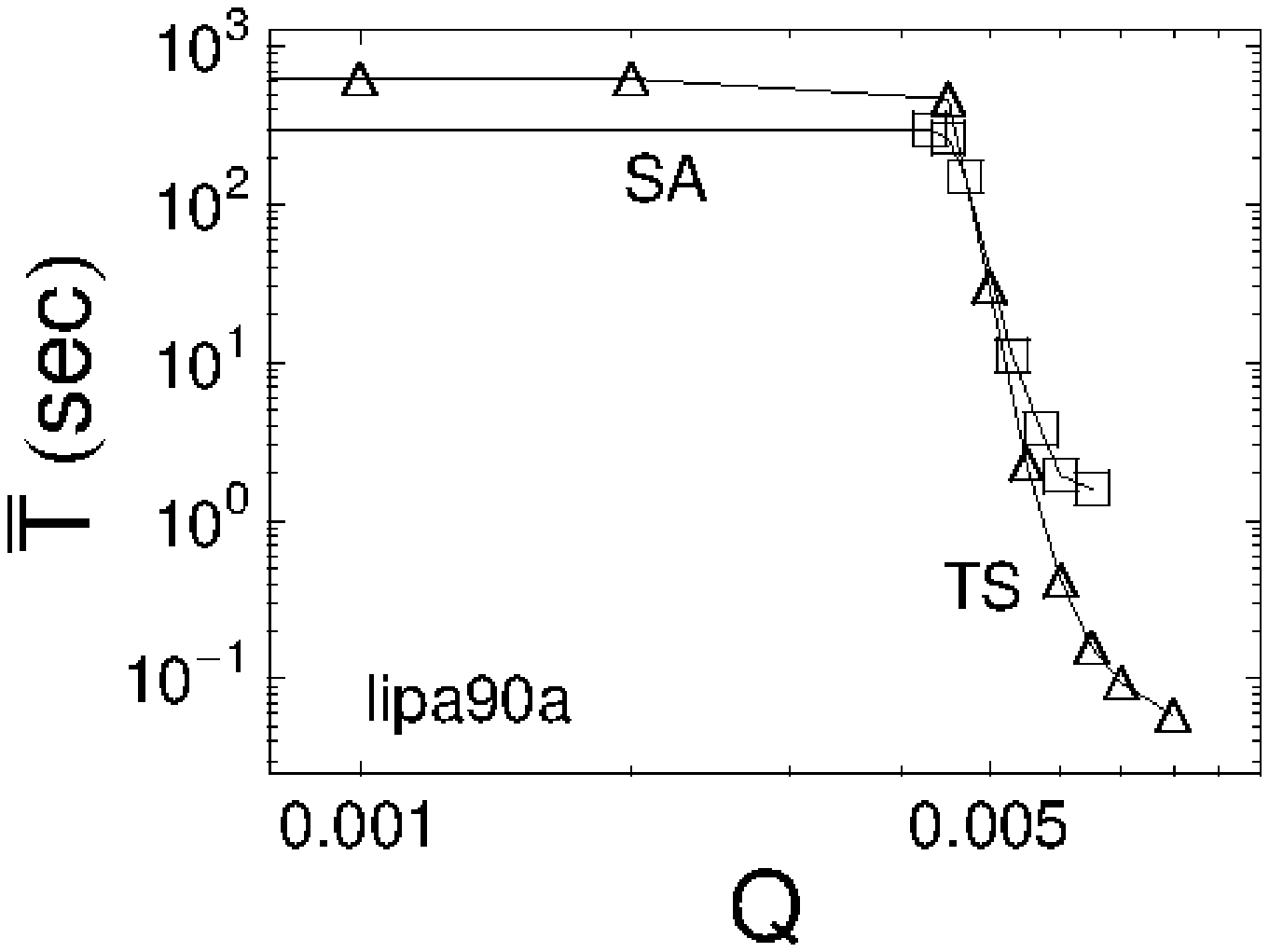}\\
\epsfxsize=7.0cm
\epsfbox{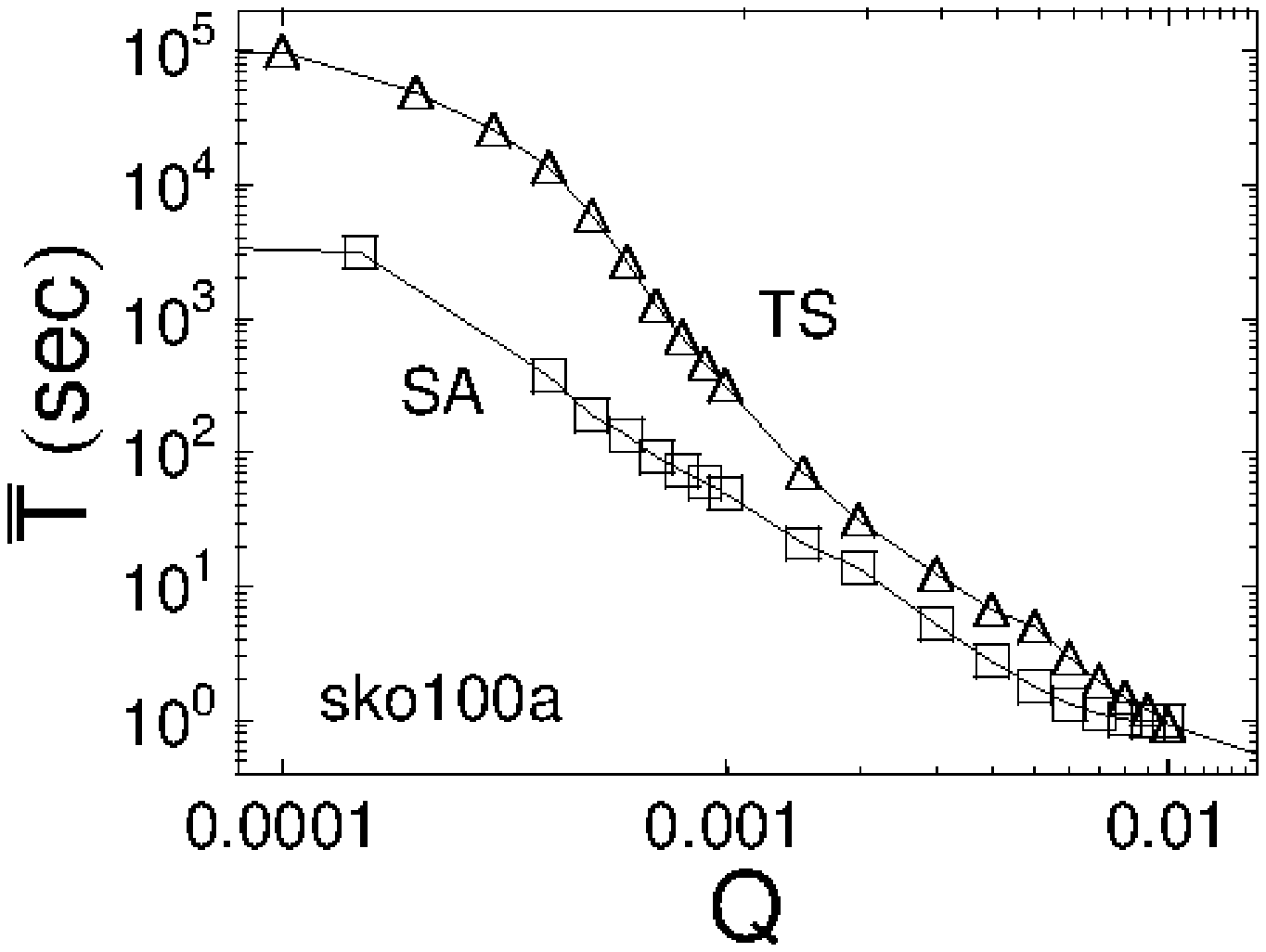} &
\epsfxsize=7.0cm
\epsfbox{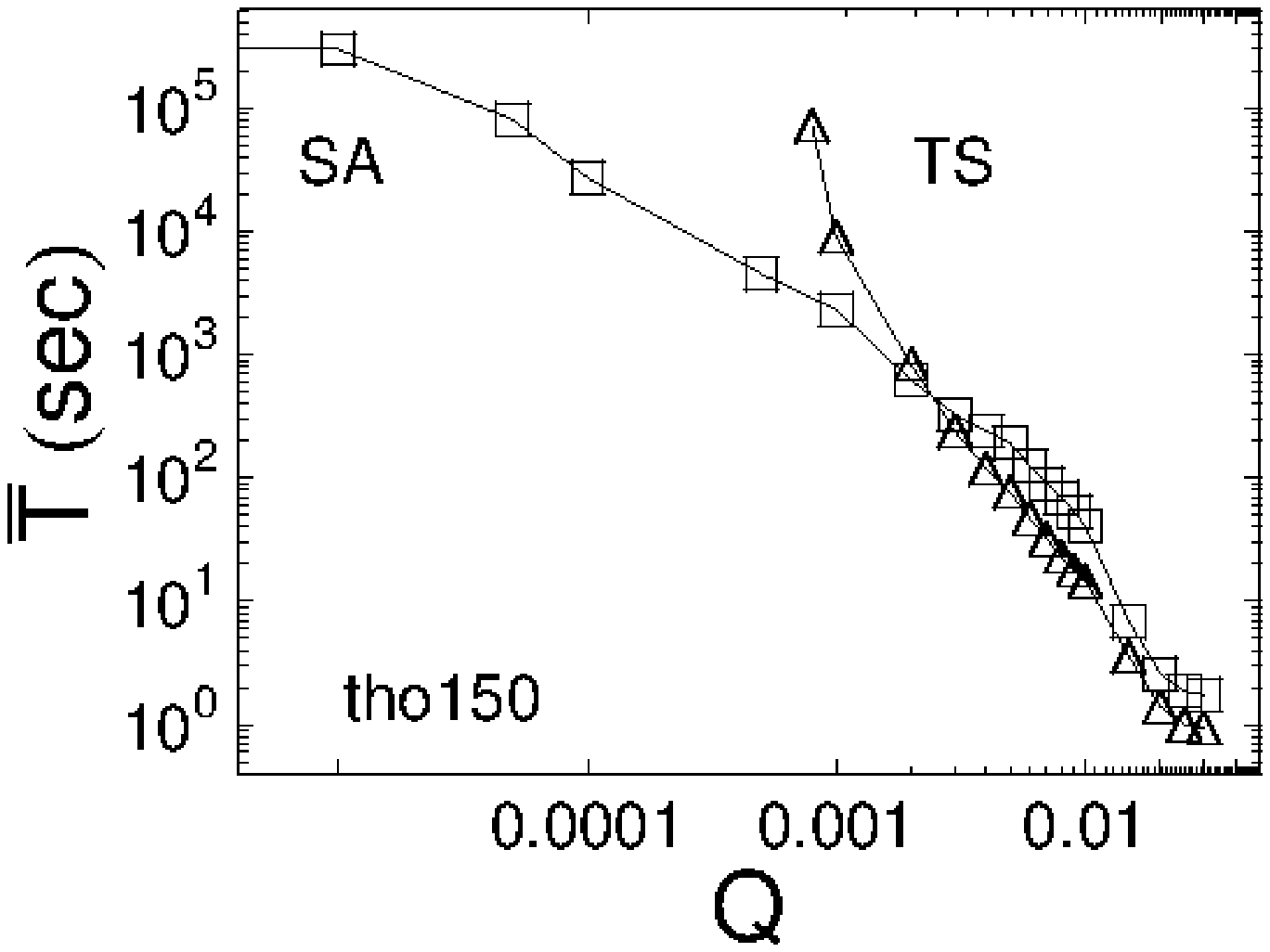}\\
\epsfxsize=7.0cm
\epsfbox{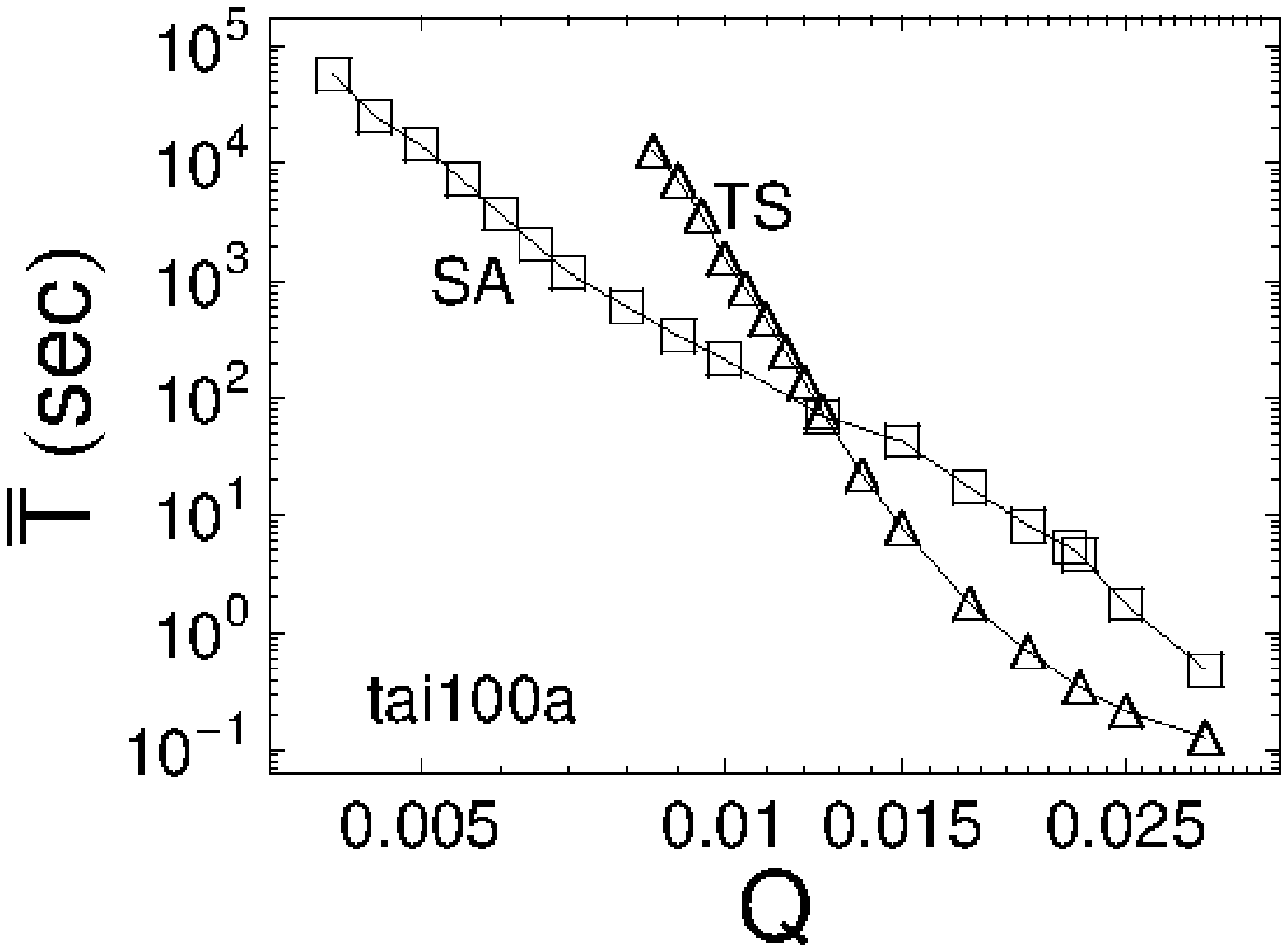} &
\epsfxsize=7.0cm
\epsfbox{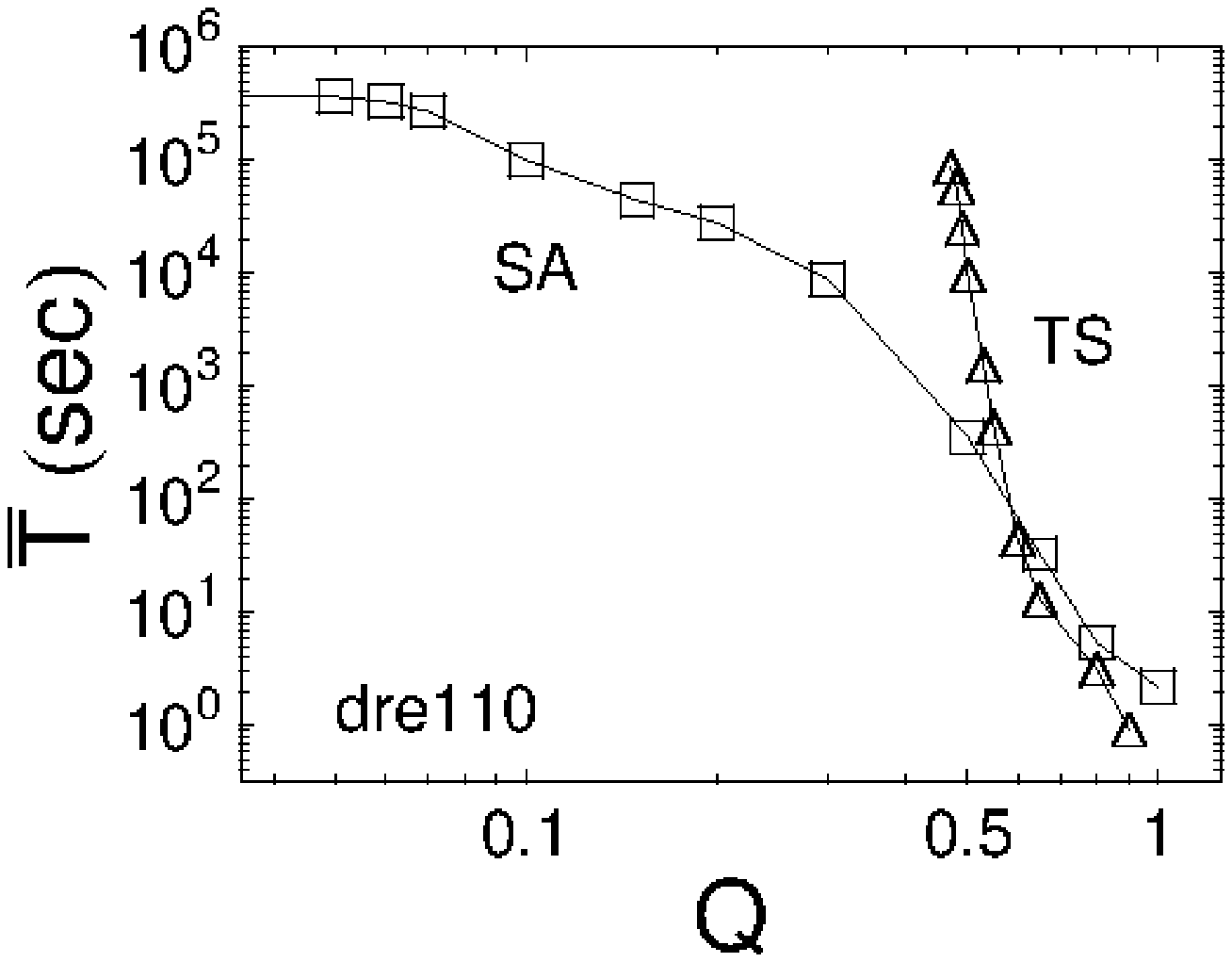}
\end{array}$
\end{center}
\caption{$\bar T$ versus Q for various problem instances for SA(squares) and TS (triangles).  For plots which 
achieve the lowest known cost for an instance ($Q=0$), we extend the
 line connecting the plot points to the left edge of the panel.   }
\label{pCRo}
\end{figure}

By plotting $\bar T(Q)$ for each heuristic we can compare the
performance of the heuristics when they are run with the optimal
number of iterations.  Fig. \ref{pCRo} plots $\bar T(Q)$ versus $Q$ for the instances studied.
Despite differences in
details, they all share the characteristic that for each problem
instance, there is a value of $Q$, $Q^*$, above which (lower quality)
tabu search performs better --- requires less time --- than simulated
annealing and below which (higher quality) simulated annealing
performs better.  SA achieves lowest known costs for all but the Tai100a instance.  
TS achieves the lowest known cost for three of the six instances.

In Table \ref{table1} we list the values of $Q^*$ for each of the
instances studied.  Note that if only the value $Q=0.01$ were
considered, the conclusion would be simply that SA is better for some
instances and TS for others.  This explains why earlier studies of
relative performance where not able to draw clear conclusions.

\subsection {Hardness of Problem Instances}

\begin{figure}[tbh]
\centerline{
\epsfxsize=7.0cm
\epsfbox{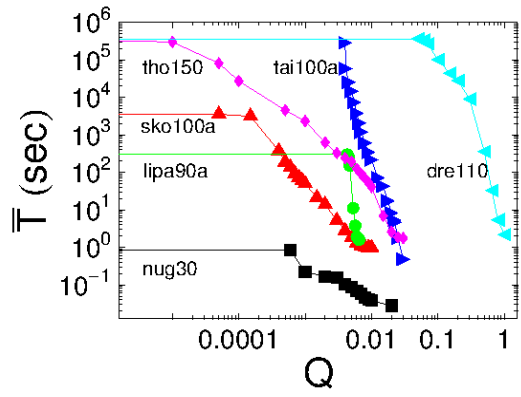}
}
\caption{{$\bar T$ versus $Q$ for for all problem instances studied: nug30 (squares), lipa90a (disks), sko100a (up-pointing triangles),
tho150 (diamonds), tai100a (right-pointing triangles, dre110 (left-pointing triangles).}
 }
\label{pAll}
\end{figure}

To compare the relative {\it hardness} of the problem instances
studied, in Fig. \ref{pAll} we plot $\bar T$ versus $Q$ for the
problem instances in a single panel.  The relative hardness of the
instances for a given solution quality is given by the relative value
of $\bar T$ at that quality. Comparing this figure with Table
\ref{table1} note that $Q^*$ appears to be correlated with the hardness
of the problem.  With the exception of Tai100a, the harder the
problem, the higher the value of $Q^*$ and thus the wider the range of
$Q$ in which SA performs better than TS.

\section{Discussion}

 How do we explain our results that, for each problem
instance studied, there is a value of the quality $Q$,
$Q^*$, above which TS performs better than SA and below which SA performs better?  A
possible qualitative explanation is that TS
essentially uses a steepest descent method to quickly find an initial
local minimum while SA finds the local minima in a more random way ---
sometimes making moves which increase the total cost even when moves
which reduce the cost could be made first. Hence for high $Q$, TS
performs better. Once a local minimum is found, however, SA is better
able to escape and find a lower minimum. As opposed to TS,
to attain better solution quality it is always better to run fewer SA
runs with a higher number of iterations.

Areas for future research might address the following questions:

\begin{itemize}
\item Is similar behavior observed when comparing SA and TS applied
  to other combinatorially complex problems?
\item When optimal numbers of iterations are used for SA and TS within
  such hybrid heuristics as hybrid genetic search, is the performance
  of the hybrid heuristic improved?
\item How does the performance of other heuristics (e.g.
  hybrid, iterated, ANT) compare when taking solution quality into account?
\item How are our findings changed if variants of TS are used? Can SA be modified to also outperform TS at high
  values of $Q$?
\end{itemize}

\section*{Acknowledgments}
We thank Jia Shao and Eric Lascaris for helpful discussions and the Defense Threat Reduction Agency (DTRA) for support.

\newpage


\begin{table}
  \caption{Value of $Q$, $Q^*$ below which SA performs better than TS.}
\centering
  \begin{tabular}{| l | r | l | }
\hline
  Problem & $C_{best}$& $Q^*$  \\
\hline

nug30 &6124 \cite{ans} & 0.0014  \\

lipa90a & 360630 \cite{lipa} & 0.0047  \\

sko100a & 152002 \cite{fleur}& 0.01  \\

tai100a &21052466 \cite{mis}& 0.0125  \\

tho150 & 8133398 \cite{mis03}& 0.025  \\

dre110 & 2052 \cite{dre}& 0.058  \\
  
\hline
\end{tabular}

\label{table1}
\end{table}

\onecolumn

\end{document}